\documentclass[prl,twocolumn,floatfix,showpacs,preprintnumbers,superscriptaddress,amsmath,amssymb]{revtex4}

\usepackage{graphicx}
\usepackage{hyperref}

\newcommand{\fig}[1]{Figure~(\ref{fig:#1})} 
\newcommand{\tbl}[1]{Table~(\ref{tbl:#1})} 
\newcommand{\cbcb}{C$_\beta$-C$_\beta$ } 

\newcommand{\tbldecoy}{
\begin{table*}[t]
\begin{center}
\begin{tabular}{|l|r|l|l|}
\hline
Name& 
RMSB& 
Energy& 
\textsf{Secondary Structure Content} \\
\hline
 N&   0.00 &         & \texttt{ccHHHHHHHHHclcbHHHHHHHHHHclcccHHHHHHHHHc} \\ \hline
 D01&  2.34 & -119.54 & \texttt{cHHHHHHHHHHHlcbcHHHHHHHHHHHHbHHHHHHHHHHc} \\ \hline
 D02&  2.41 & -117.52 & \texttt{cHHHHHHHHHHHlcbHHHHHHHHHHHHHbHHHHHHHHHHc} \\ \hline
 D03&  2.76 & -116.25 & \texttt{cHHHHHHHHHHHlcbHHHHHHHHHHHHHbHHHHHHHHHHc} \\ \hline
 D04&  2.40 & -115.85 & \texttt{cHHHHHHHHHHHlbbHHHHHHHHHHHHHbHHHHHHHHHHc} \\ \hline
 D05&  2.43 & -114.67 & \texttt{cHHHHHHHHHHHlcbHHHHHHHHHHHcbHHHHHHHHHHHc} \\ \hline
 D06&  6.48 & -114.06 & \texttt{cHHHHHHHHHHHcccbHHHHHHHHHHHHbHHHHHHHHHHc} \\ \hline
 D07&  2.57 & -113.65 & \texttt{cHHHHHHHHHHHlbbcHHHHHHHHHHHHbHHHHHHHHHHc} \\ \hline
 D08&  4.61 & -107.72 & \texttt{cHHHHHHHHHcclccHHHHHHHHHHHHHlclHHHHHHHHc} \\ \hline
 D09&  4.14 & -106.29 & \texttt{cHHHHHHHHHHHcbcbHHHHHHHHHbblcHHHHHHHHHHc} \\ \hline
 D10&  5.92 & -103.88 & \texttt{cHHHHHHHHHHHlcHHHHHHHHHbcbcclbHHHHHHHHHc} \\ \hline
\end{tabular}
\caption{
Table of the 10 lowest energy decoys (of 20, remainder available by
request from the authors) with backbone RMS deviation to the NMR structure and 
secondary structure content. The first row designates the secondary 
structure content of the NMR structure.}
\label{tbl:decoys}
\end{center}
\end{table*}

}

\begin{document}
\title{{\em In-silico} folding of a three helix protein and
characterization of its free-energy landscape in an all-atom forcefield}

\author {T. Herges} 
\affiliation{Forschungszentrum Karlsruhe, Institut f\"ur Nanotechnologie,
76021 Karlsruhe, Germany}

\author {W. Wenzel} 
\affiliation{Forschungszentrum Karlsruhe, Institut f\"ur Nanotechnologie,
76021 Karlsruhe, Germany}

\date{June 18, 2003}

\begin{abstract}
We report the reproducible first-principles folding of the 40 amino
acid, three-helix headpiece of the HIV accessory protein in a recently
developed all-atom free-energy forcefield. Six of twenty simulations
using an adapted basin-hopping method converged to better than 3 \AA\
backbone RMS deviation to the experimental structure. Using over
60,000 low-energy conformations of this protein, we constructed a
decoy tree that completely characterizes its folding funnel.
\end{abstract}

\pacs{87.15.Cc,02.70.Ns,02.60.Pn}

\maketitle

Available genomic and sequence information for proteins contains a
wealth of biomedical information that becomes accessible when
translated into three-dimensional structure~\cite{baker01}. While
theoretical models for protein structure
prediction~\cite{schonbrunn02,liwo02} that partially rely on
experimental information have shown consistent progress\cite{moult01},
the assessment of de-novo strategies that rely on sequence information
alone has been much less favorable~\cite{bonneau01}. The development
of such techniques, in particular of transferable first-principle
all-atom folding methods, would significantly benefit the
understanding of protein families where little experimental
information is available, the prediction of novel folds as well as the
investigation of protein association and dynamics which are presently
difficult to probe experimentally. Recent progress for small
peptides~\cite{liwo02,snow02,simmerling02,schug03} documents both the
feasibility of this approach as well as its
limitations~\cite{hansmann02,lin03}, in particualr those associated with the
direct simulation of the folding process through molecular
dynamics~\cite{duan98}.

Overwhelming experimental evidence supports the thermodynamic
hypothesis~\cite{anfinsen73} that many proteins are in thermodynamic
equilibrium with their environment: their native state thus
corresponds to the global minimum of their free energy
landscape~\cite{onuchic97}. The free energy of the system is
accessible either indirectly by explicit ensemble averaging of the
combined internal energy of protein and solvent, or directly in a
free-energy forcefield where an implicit solvation model approximates
direct interactions with the solvent as well as most of the entropic
contributions. We developed an all-atom protein forcefield
(PFF01)~\cite{herges03b,schug03,herges04a} with an area-based implicit
solvent model that approximates the free energy of peptide
conformations in the natural solvent. Using a rational decoy approach
this forcefield was optimized to correctly predict the native
structure of the 36-amino acid headgroup of
villin~\cite{duan98,hansmann02,lin03}. Without further parameter
adjustment we then simulated the structurally conserved 40 amino-acid
headpiece of the autonomously folding HIV accessory protein
(1F4I-40)~\cite{withers00} with a modified basin hopping
technique~\cite{nayeem91,wales96}. Out of twenty simulations the five
energetically lowest correctly reproduced the NMR structure of this
three-helix protein with a backbone RMS deviation of less than 3
{\AA}. The combination of decoy based model development for the free
energy with efficient stochastic optimization methods suggests a
viable route for protein structure prediction at the all atom level
with present day computational resources.

{\em Model:} We have recently developed an all-atom protein
forcefield (PFF01), which was used to reproducibly fold the 20 amino
acid trp-cage protein~\cite{schug03}. 
PFF01~\cite{herges04a} comprises an atomically resolved electrostatic
model with group specific dielectric constants and a Lennard Jones
parameterization that was adapted to the experimental distance
distributions from crystal structures of 138
proteins~\cite{abagyan94}. Interaction with the fictitious solvent are
modeled in a simple solvent accessible surface
approach~\cite{eisenberg86}, where the solvation free-energies per
unit surface were fitted to the enthalpies of solvation of the
Gly-X-Gly series of peptides~\cite{sharp91}. The only low-energy degrees of
freedom available to the peptide during the folding process are
rotations of the dihedral angles of the backbone and the sidechains,
these are the only moves considered during the simulation. There are
two move-classes, small random rotations about a single angle and
library moves, which set a particular backbone dihedral to a permitted
value in the Ramachandran plot.

\tbldecoy

If an accurate model for the free energy of the protein in its
environment is available, stochastic optimization methods can be used
to locate the global optimum of the free-energy landscape orders of
magnitude faster than traditional simulation techniques. We adapted
the basin hopping technique~\cite{nayeem91} (BHT) for protein
simulations by replacing a single minimization step with a simulated
annealing run~\cite{kirkpatrick83} with self-adapting cooling cycle
and length. At the end of one annealing step the new conformation was
accepted if its energy difference to the current configuration was no
higher than a given threshold energy $\varepsilon _{T}$, an approach
recently proven optimal for certain optimization
problems~\cite{schneider98}.  Each simulation was performed in three
separate steps: First we used a high temperature bracket of 800/300 K
($\varepsilon _{T}$ =15 K) for the annealing window and reduced the
strength of the solvent terms in the forcefield by 20{\%}. The second
step started from the final configurations of the first run, used the
same annealing window but the full solvent interactions. In the third
step the resulting structures were further annealed in a low
temperature bracket of 600/3 K ($\varepsilon _{T}$=1K). Within each
annealing run the temperature was geometrically decreased, also the
number of steps per annealing run was gradually increased to ensure
better convergence. In total each simulation comprised 10$^{7}$ energy
evaluations with a computational effort roughly corresponding to a 10
ns MD simulation (in vacuum with 1 fs timestep). After this time no
significant energy fluctuations occurred in the simulations,
indicating that each had settled into a metastable configuration.

{\em Results:} Using PFF01 we performed 20 independent modified basin
hopping simulations of the structurally conserved 40 amino-acid
headpiece of the HIV accessory protein (pdb-code 1F4I, sequence:
QEKEAIERLK ALGFEESLVI QAYFACEKNE NLAANFLLSQ). The best structures
found in each run were ranked according to their energy and the
backbone RMS deviation (RMSB) to the NMR structure was
computed. \tbl{decoys} demonstrates that the five lowest structures
had to good accuracy converged to the NMR structure of the
protein. The first non-native decoy appears in position six, with an
energy deviation of 5 kcal/mol (in our model) and a significant RMSB
deviation. The table demonstrates that all low-energy structures have
essentially the same secondary structure, i.e. position and length of
the helices are always correctly predicted, even if the protein did
not fold correctly. The degree of secondary structure content and
similarity decreases for the decoys with higher energy (data not
shown) in good correlation with their energy. From the standpoint of
the optimization approach (but not necessarily for the physical
folding scenario) this suggests that successful folding requires the
formation of the correct secondary structure content as a
prerequisite.

\begin{figure}[bp]
\centerline{\includegraphics[width=0.9\columnwidth]{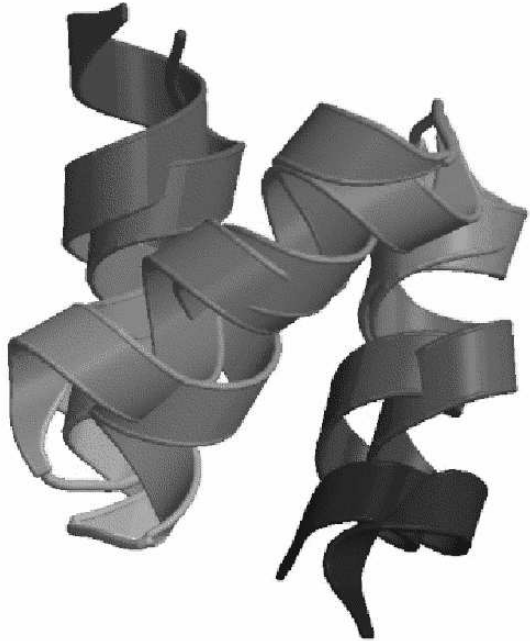}}
\caption{Overlay of the secondary structure elements of the native 
configuration and the folded structure of 1F4I-40}
\label{fig:overlay}
\end{figure}
\begin{figure}[t]
\centerline{\includegraphics[width=0.7\columnwidth]{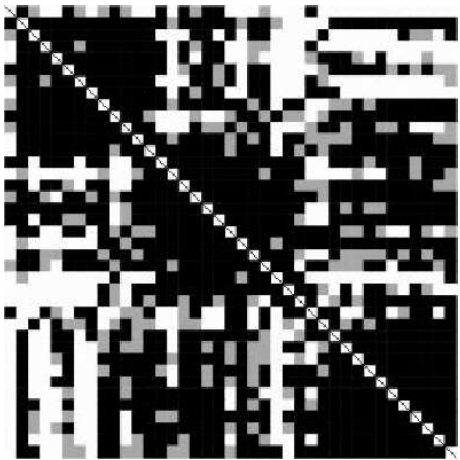}}
\caption{ Color coded \cbcb distance error map for the
folded structure 1F4I in comparison to the NMR structure. Each square
encodes the deviation between the \cbcb distance of two
amino acids in the NMR to the C$\beta $-C$\beta $ distance of the same
amino acids in folded structure. Black (grey) squares indicate a
deviation of less than 1.50 {\AA} (2.25 {\AA}). White squares indicate
large deviations.}
\label{fig:cbcb}
\end{figure}

The good agreement between the folded and the experimental structure
is also evident from \fig{overlay}, which shows the secondary
structure alignment of the native and the folded conformations. The
good physical alignment of the helices illustrates the importance of
hydrophobic contacts to correctly fold this protein. An independent
measure to assess the quality of these contacts is to compare the
\cbcb distances (which correspond to the NOE constraints of the NMR
experiments that determine tertiary structure) in the folded structure
to those of the native structure. The color coded \cbcb distance in
\fig{cbcb} demonstrates a 66 \% (80 {\%}) coincidence of the \cbcb distance
distances to within one (1.5) standard deviations of the experimental
resolution. The dark diagonal block indicate intra-helical contacts,
which are, perhaps not too surprisingly, resolved to very good
accuracy. The off-diagonal dark blocks, however, indicate that also
a large fraction of long range native contacts is reproduced
correctly. 

\begin{figure}[b]
\centerline{\includegraphics[width=\columnwidth]{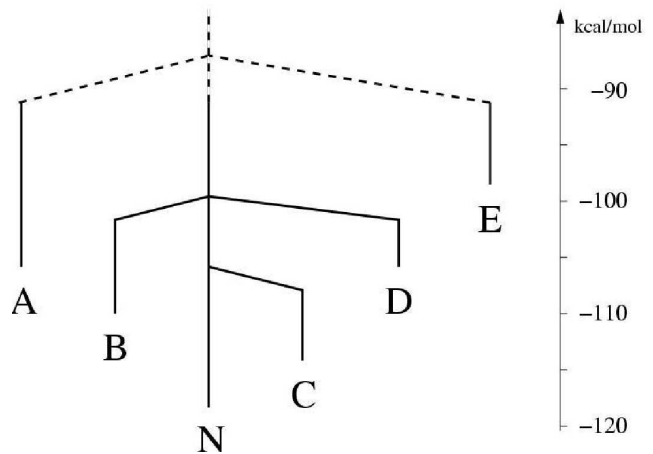}}
\caption{Decoy tree of the low energy configurations of the 40 amino
  acid headpiece of th HIV accessory protein. The horizontal axis
  depicts the total energy the chart on the right the total number of
  decoys in the primary and secondary funnels. }
\label{fig:tree}
\end{figure}

Starting from intermediate structures of the folding simulations we
generated over 60,000 low-energy conformations (decoys). Decoys with a
root mean square deviation of the backbone (RMSB) deviation of less
than 3 {\AA} were grouped into families with free energy brackets of 2
kcal/mol. We then resolved the topological
hierarchy\cite{wales96,brooks01} of the associated potential energy
surface through the construction of a decoy tree (\fig{tree}) that
illustrates the low-energy structure of the free energy surface.
Beginning from the best conformation, we draw a vertical line for each
decoy family in this window. Moving upward in energy the number of
decoys in each family grows almost exponentially in the low energy
region which we can resolve well. As a result the diversity of each
family grows until different families unite. Family membership is
associative, i.e. as soon as two decoys in different branches have an
RMSB deviation of less than 3 \AA, all members of both families belong
to one superfamily. Pictorially this representation results in an
inverted tree-like structure that characterizes the topology of the
metastable states of the free energy surface.

Trees with very short stems and many low-energy branches are
characteristic of glassy potential energy surfaces, which are
associated with Levinthals paradox~\cite{levinthal68,honig99} in the
context of protein folding. Well structured trees with few terminal
branches suggest the existence of a folding funnel~\cite{onuchic97},
consistent with the ``new'' paradigm for protein
folding~\cite{karplus94}. From this perspective the structure in
\fig{tree} this tree is indicative of the existence of a very broad
folding funnel~\cite{onuchic97} with pronounced competing secondary
metastable conformations, which are depicted as the non-native
terminal minima of the tree. The discretization of the energy axis in
intervals of 2 kcal/mol starting from the native conformation results
in a smoothing of the free-energy surface. Each line in the figure
corresponds to a family containing hundreds tothousands of structures,
which are all associated with the same low-lying metastable
conformation (the terminal point of the branch). Simulations trapped
in such a metastable state must overcome a potential energy barrier of
the order of the energy difference to the next highest branching point
of the tree to visit another structure. The branching points of the
tree were constructed only from structures of the decoy set and not
through independent transition state search among the terminal
structures. In addition, main-chain entropy is neglected in this
analysis, which results in an overestimation of the barrier. Further
investigations to more accurately determine the transition states are
presently under way.

\begin{figure}[t]
\centerline{\includegraphics[width=0.8\columnwidth]{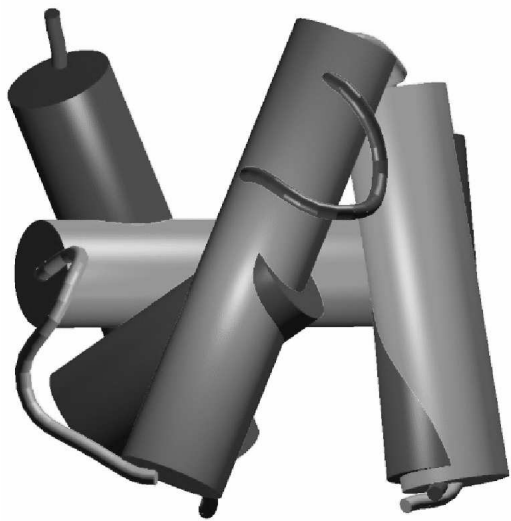}}
\caption{Overlays of the secondary structure elements of the native (green)
configuration and the lowest misfolded decoy (red) of 1F4I.}
\label{fig:misfold}
\end{figure}

The lowest competing terminal branch (branch C), associated with decoy
D06 in \tbl{decoys}, is less than 5 kcal/mol above the best native
decoy. Decoy D06 has comparable energy to competing decoys but much
larger RMSB deviation and has few long-range native contacts. This
structure (see \fig{misfold}) has also three helices of comparable
length and sequence location, but differs from the native structure in
the relative alignment of the helices with respect to each. The RMSB
deviation of the low-lying terminal structures to the NMR structures
is large (i.e. comparable with the RMSB deviation to unfolded
structures), indicating that conserved secondary structure elements,
rather than distance constraints characterize the folding funnel. The
low lying local minima are thus characterized by varying spatial
arrangements of similar secondary structure elements, a property that
is ill represented by either RMSB deviation or the number of native
contacts.

{\em Discussion:} With this work we have demonstrated that accurate
free-energy forcefields can predict the native structure of proteins
with nontrivial tertiary structure with present day computational
resources. This result represents an {\em in-silico} realization of
the thirty year-old thermodynamic hypothesis that proteins are in
thermodynamic equilibrium with the environment under physiological
conditions~\cite{anfinsen73}. Under this hypothesis, the native
structure of the protein can be predicted using stochastic
optimization methods orders of magnitude faster than by direct
simulation. Our results demonstrate that the important influence of
the solvent can be modeled with a relatively simple solvent accessible
surface approach. 

The analysis of the free energy surface supports the funnel paradigm
of protein folding for a nontrivial protein with a significantly
larger hydrophobic core than was previously possible. The relatively
small number of terminal branches of the decoy tree offers the first
glimpse on the experimentally inaccesible structure of the folding
funnel. It suggest the exsitence of a broad folding funnel with well
defined secondary metastable states which may constitute important
folding intermediates. The free energy optimization approach used here
permitted the characterization of these low-lying states, which
surprisingly share very similar secondary structure with the native
configurations. Investigations of other proteins must show, whether
this pattern persist also for other proteins.

It should be noted that the success of the optimization approach
depends strongly on the ability of the optimization technique to
differentiate between the low-lying minima of the FES {\em in a
realistic forcefield}. The performance of optimization methods is
often strongly problem dependent, but with 1F4I, 1VII and 1L2Y three
nontrivial model systems exist on which different optimization methods
can be evaluated.  The decoy sets generated and insights regarding
low-lying metastable states can also serve as a test-bed for the
development of coarse-grained protein models. PFF01 is presently be
validated for other peptides and proteins and rationally evolved to
correctly predict the structure of larger fold families.

Acknowledgment: We are grateful to S. Gregurick, J. Moult and J. Pedersen 
for discussions and portions of the code used in these simulations. This 
work was supported by the Deutsche Forschungsgemeinschaft, the Bode 
Foundation and the BMWF.

\bibliographystyle{apsrev}
\bibliography{optimization,protein,docking}
\end{document}